\documentclass[a4paper]{article}

\usepackage[english]{babel}
\usepackage[utf8x]{inputenc}
\usepackage[T1]{fontenc}

\usepackage[a4paper,top=3cm,bottom=2cm,left=3cm,right=3cm,marginparwidth=1.75cm]{geometry}

\usepackage{setspace}
\usepackage{amssymb}

\usepackage{cite}

\usepackage{amsmath}
\usepackage{graphicx}
\usepackage[colorinlistoftodos]{todonotes}
\usepackage[colorlinks=true, allcolors=blue]{hyperref}

\title{Nonvolatile, reconfigurable and narrowband mid-infrared filter based on surface lattice resonance in phase-change GeSbTe}

\author{Xingzhe Shi$^{1,2}$, Changshui Chen$^{1}$, Songhao Liu$^{1}$, and Guangyuan Li$^{2,3,*}$}

\date{}
\begin{document}
\maketitle

\begin{spacing}{2.0}

\noindent \large$^1$Guangdong Provincial Key Laboratory of Nanophotonic Functional Materials and Devices, School of Information Optoelectronic Science and Engineering, South China Normal University, Guangzhou, 510006, China

\noindent  $^2$CAS Key Laboratory of Human-Machine Intelligence-Synergy Systems, Shenzhen Institutes of Advanced Technology, Chinese Academy of Sciences, Shenzhen 518055, Guangdong Province, China

\noindent  $^3$Shenzhen College of Advanced Technology, University of Chinese Academy of Sciences, Shenzhen 518055, China


\noindent *Corresponding author: gy.li@siat.ac.cn

\end{spacing}

\begin{abstract}
We propose a nonvolatile, reconfigurable, and narrowband mid-infrared bandpass filter based on surface lattice resonance in phase-change material Ge$_2$Sb$_2$Te$_5$ (GST). The proposed filter is composed of a two-dimensional gold nanorod array embedded in a thick GST film. Results show that when GST transits from the amorphous state to the crystalline state, the narrowband reflection spectrum of the proposed filter is tuned from $3.197~\mu$m to $4.795~\mu$m, covering the majority of the mid-infrared regime, the peak reflectance decreases from 72.6\% to 25.8\%, and the corresponding Q-factor decreases from 19.6 to 10.3. We show that the spectral tuning range can be adjusted by varying the incidence angle or the lattice period. By properly designing the gold nanorod sizes, we also show that the Q-factor can be greatly increased to 70 at the cost of relatively smaller peak reflection efficiencies, and that the peak reflection efficiency can be further increased to 80\% at the cost of relatively smaller Q-factors. We expect this work will advance the engineering of GST-based nonvalatile tunable surface lattice resonances and will promote their applications especially in reconfigurable narrowband filters.
\end{abstract}

\section{Introduction}
Dynamically tunable narrowband mid-infrared (3~$\mu$m to 5~$\mu$m) filters are key devices in a diverse range of applications, including chemical spectroscopy, thermography, multispectral/hyperspectral imaging \cite{OLE2006Monitoring,ApplSpec2012Spec,CBP2016thermography}. It has been accepted that conventional tunable narrowband mid-infrared filters realized by using diffraction gratings, motorized filter wheels, acousto-optic interactions \cite{ProcSPIE2018MidIRfilterAO,AO2018MidIRfilterAO}, or Fabry-P\'erot interferometers based on micro-electro-mechanical systems (MEMS) \cite{ProcSPIE2016MidIRfilterMEMS} or liquid-crystals \cite{JMM2020MidIRfilterLC} suffer from different limitations, as summarized by Julian {\sl et al.} \cite{Optica2020GstFilterEOT}. For example, diffraction gratings and motorized filter wheels have moving parts and slow tuning speed, acousto-optic tunable filters are complex to manufacture and integrate, and Fabry-P\'erot interferometer-based filters have limited spectral tunability.

Recently, chalcogenide phase-change materials, especially germanium-antimony-telluride (Ge$_x$Sb$_y$Te$_z$, GST), have attracted increasing attention because of their appealing merits such as nonvolatile, rapid and reversible switching between the amorphous and the crystalline states by electrical or short optical pulses, tremendous differences in optical and electronic properties between these two states, and high chemical and long-term stability \cite{NatPhoton2017PCMrev,ATS2019PCMrev,AOM2019PCMmsRev,NanoP2020PCMrev}. Based on GST as a nonvolatile reconfigurable platform, various tunable mid-infrared filters have been proposed or demonstrated. Cao {\sl et al.} \cite{JOSAB2013GstAbsTuneMIM,JOSAB2013GstAbsTuneMIM2}, Tittl {\sl et al.} \cite{AM2015GstAbsTuneMIM}, and Tian {\sl et al.} \cite{EPL2019GstAbsTuneMIM} respectively proposed or demonstrated tunable mid-infrared filters based on perfect absorbers that are composed of one- or two-dimensional metal metasurface on top of a metallic mirror sandwiched by a GST film. However, only bandstop (notch) filters in reflection mode can be realized based on perfect absorbers, and the spectral tunability is limited (only $\sim300$~nm in \cite{JOSAB2013GstAbsTuneMIM}, $\sim500$~nm in  \cite{AM2015GstAbsTuneMIM}, $1300$~nm in \cite{JOSAB2013GstAbsTuneMIM2}, or $1464$~nm in \cite{EPL2019GstAbsTuneMIM}). Ding {\sl et al.} \cite{SM2019GstFilterRg} also proposed tunable bandstop mid-infrared filters based on GST metasurface composed of nanorods. In order to achieve bandpass filters, Zhou {\sl et al.} \cite{AO2020GstFilterFP} and Williams {\sl et al.} \cite{OE2020GstFilterFP} respectively proposed or demonstrated tunable mid-infrared bandpass filters by embedding a GST film in a Fabry-P\'erot cavity. Although the Q-factor, defined as the ratio of the full-width-half-maximum (FWHM) to the central wavelength, can reach 70--90, the spectral tunability is very limited (only $\sim420$~nm in \cite{AO2020GstFilterFP} or $\sim300$~nm in \cite{OE2020GstFilterFP}). Rud\'e {\sl et al.} \cite{AOM2016GstFilterEOT} combined GST thin films with extraordinary optical transmission (EOT) through periodic arrays of subwavelength nanoholes drilled in a thick gold film, and demonstrated that the resonant wavelength can shift by 385~nm. Later Trimby {\sl et al.} \cite{SPIE2018GstFilterEOT} discussed approaches to modify the filter performance, including the Q-factor, the spectral range, and the peak transmission. Quite recently, Julian {\sl et al.} \cite{Optica2020GstFilterEOT}  demonstrated reversible mid-infrared filters with high transmittance ($\sim70\%$) and narrowband performance ($Q\sim45$). However, the spectral tunability (500~nm, from 2.91~$\mu$m to 3.41~$\mu$m) is still too small to cover the mid-infrared regime.


Recently, plasmonic surface lattice resonances (SLRs), which are collective Fano resonances formed by the diffraction coupling of localized surface plasmon resonance \cite{ChemRev2018GrigorenkoSLRreview,NanoRes2018ZhengSLRreview,MatToday2018OdomSLRreview}, have been of particular interest in ultra-narrowband absorbing/filtering applications \cite{ACSNano2014SLRfilter}. In 2013, Chen {\sl et al.} \cite{OE2013GstFilterSLR} demonstrated nonvolatile tuning of SLRs over a range of $\sim$500~nm in the near-infrared regime (1.89~$\mu$m to 2.27~$\mu$m) by incorporating a thin GST film between a gold nanodisk array and a quartz substrate. However, the obtained transmission spectra profiles are not suitable for narrowband filtering applications. Michel {\sl et al.} \cite{NL2013GstFilterSLR} demonstrated tunable filters by combining thin GST films with nanoantennas that support SLRs, and achieved resonant wavelength shift up to 668~nm (from 3.926~$\mu$m to 4.594~$\mu$m) when the GST phase transits from the amorphous state to the crystalline state. However, the reflectance was not provided, and the Q-factor is very low ($\sim5$) due to the inhomogeneous refractive index environment.

In this work, we propose a novel nonvolatile and reconfigurable mid-infrared bandpass filter based on tunable SLR supported by a two-dimensional (2D) array of gold nanorods, which are embedded in a thick Ge$_2$Sb$_2$Te$_5$ film. The operation principle will be elaborated. Results will show that the reflection spectra of the proposed structure can be dynamically tuned by changing the GST crystallization fraction, which can be achieved with a single nanosecond laser pulse \cite{Optica2020GstFilterEOT}. Remarkably, we will show that our design has extremely large spectral tunability of 1.598~$\mu$m (from 3.197~$\mu$m to 4.795~$\mu$m), which covers the majority of the mid-infrared regime, high reflection efficiencies and relatively large Q-factors ($R=72.6\%$ and $Q=19.6$ at 3.197~$\mu$m, $R=25.8\%$ and $Q=10.6$ at 4.795~$\mu$m). The underlying physics will be clarified with near-field distributions as well as the strong dependence of the reflection spectra on the lattice period and the incidence angle. By investigating the effects of gold nanorod sizes, we will also show that the filtering performance can be further adjusted to achieve even larger Q-factor or peak reflectance.

\section{Theory and simulation setup}
\begin{figure}[htbp]
\centering
\includegraphics[width=0.98\linewidth]{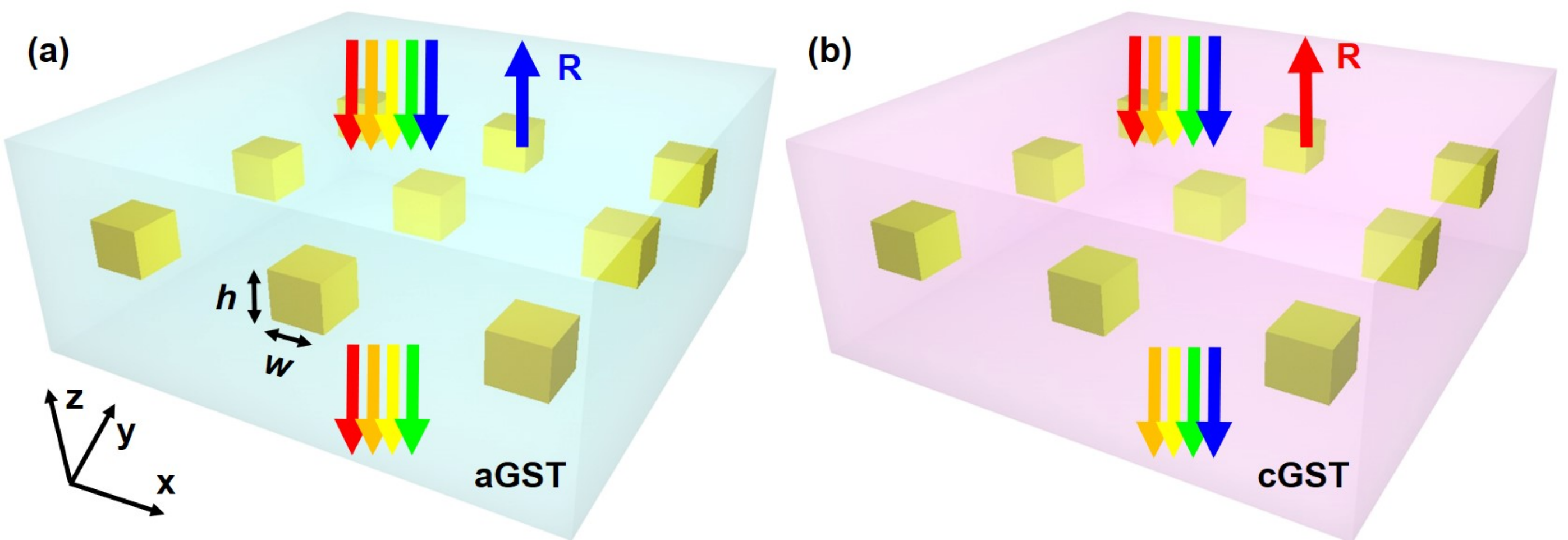}
\caption{Schematic of the proposed tunable mid-infrared bandpass filter composed of Au nanorod array embedded in a thick GST film. When GST transits from (a) the amorphous phase (denoted as ``aGST'') to (b) the crystalline phase (denoted as ``cGST''), the narrowband reflection spectrum experiences a large red shift.} \label{fig:schem}
\end{figure}

Figure~\ref{fig:schem} illustrates the proposed nonvolatile reconfigurable narrowband mid-infrared filter. The design is composed of a 2D array of gold nanorods that are embedded in an optically thick GST film. The square-shaped gold nanorods have side length of $w=200$~nm and thickness of $h=180$~nm, and the lattice periods along both the $x$ and $y$ axes are equal to $\Lambda$. The structure is illuminated by normally incident plane wave with unitary electric field amplitude and linear polarization along the $x$ direction. 

\begin{figure}[htbp]
\centering
\includegraphics[width=0.7\linewidth]{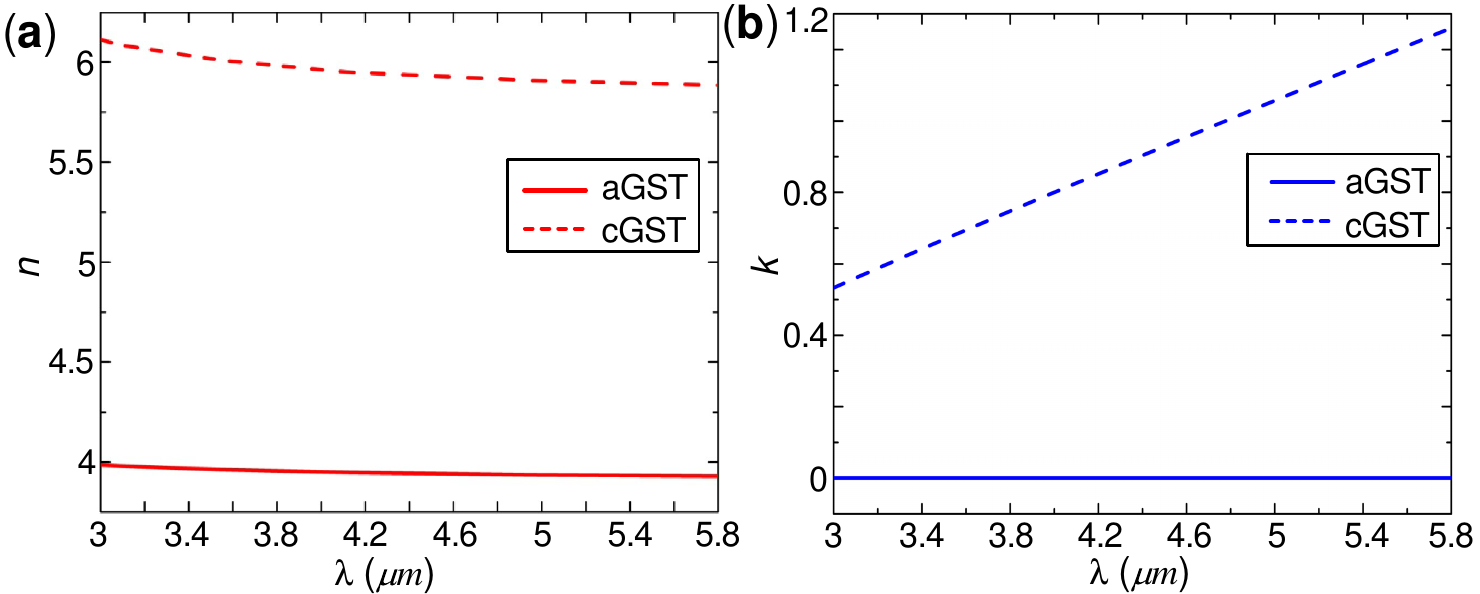}
\caption{(a) Real and (b) imaginary parts of GST's refractive index as a function of the wavelength \cite{ProcSPIE2017GSTnk}. Solid and dashed curves are for GST in the amorphous state (denoted as ``aGST'') and in the crystalline state (denoted as ``cGST''), respectively.}
\label{fig:GSTnk}
\end{figure}

The operation principle of the proposed filter is as follows. For such a gold lattice surrounded by homogeneous GST environment, the Rayleigh anomaly (RA) wavelength of the $(p,q)$ diffraction order, $\lambda_{\rm RA}^{(p,q)}$, can be calculated by
\begin{equation}
\label{Eq:RA}
k_{0}n=\left | \overrightarrow{k}_{\parallel } +\overrightarrow{G}\right | \,,
\end{equation}
where $\overrightarrow{k}_{\parallel}=k_{0}n \sin\theta$ with $k_0=2\pi/\lambda_{\rm RA}^{(p,q)}$ and $\theta$ the incidence angle, $n$ is the refractive index of GST, and $\overrightarrow{G}=\left ( p\frac{2\pi }{\Lambda}, q\frac{2\pi }{\Lambda} \right )$ is the reciprocal vector that is inversely proportional to the lattice period $\Lambda$. Thus under normal incidence ($\theta=0$) we have 
\begin{equation}
\label{Eq:RA2}
\lambda_{\rm RA}^{(p,q)}=n\Lambda/\sqrt{p^2+q^2}\,.
\end{equation}
If the gold nanorod array are well designed such that SLR can be excited at wavelength that is close to the RA wavelength of the $(\pm1, 0)$ order, $\lambda_{\rm RA}^{(\pm1,0)}=n\Lambda$, the SLR wavelength can be tuned by $\sim(n_{\rm c}-n_{\rm a})\Lambda$ when GST transits from the amorphous state with refractive index $n_{\rm a}$ to the crystalline state with refractive index $n_{\rm c}$. In the mid-infrared regime, $n_{\rm a}\approx4$ and $n_{\rm c}\approx6$ \cite{ProcSPIE2017GSTnk}, as shown by Fig.~\ref{fig:GSTnk}, thus we can expect a large spectral tunability of $\sim2\Lambda$, which is twice of the lattice period $\Lambda$. 
By taking $\Lambda = 750$~nm so that the RA wavelengths of the $(\pm1, 0)$ order are $n_{\rm a}\Lambda \approx 3~\mu$m and $n_{\rm c}\Lambda \approx 4.5~\mu$m for GST in the amorphous and the crystalline states, respectively, we are able to obtain extremely large spectral tunability reaching $\sim1.5~\mu$m, which covers the vast majority of the mid-infrared 3~$\mu$m to 5~$\mu$m atmospheric window.

All the simulations were performed with a home-developed package for fully vectorial rigorous coupled-wave analysis (RCWA) following \cite{JOSAA1995RCWA,JOSAA1997RCWA,PRB2006RCWA}. As a powerful tool for modelling periodic photonic structures, the RCWA technique can calculate the reflectance and the transmittance, as well as the near-field electric and magnetic field distributions. The numbers of the 2D Fourier harmonics in our RCWA simulations were $31\times31$, which were confirmed to be enough to reach the convergence regime. The wavelength-dependent permittivities of gold were taken from \cite{Au2012PRB}. The effective wavelength-dependent permittivity of GST in various crystallization conditions can be described by the Lorenz–Lorentz relation \cite{AJP1982LLmodel},
\begin{equation}
\label{Eq:epsiEff}
\frac{\varepsilon_{\rm eff}(\lambda)-1}{\varepsilon_{\rm eff}(\lambda)+2}=m\frac{\varepsilon _{\rm c}(\lambda)-1}{\varepsilon_{\rm c}(\lambda)+2}+\left ( 1-m \right )\frac{\varepsilon_{\rm a}(\lambda)-1}{\varepsilon_{\rm a}(\lambda)+2} \,,
\end{equation}
where $m$ is the crystalline fraction of GST, ranging from 0 to 1. $\varepsilon_{\rm a}(\lambda)$ and $\varepsilon _{\rm c}(\lambda)$ are the dielectric constants of GST in the amorphous ($m=0$) and in the crystalline ($m=1$) states, respectively, which are calculated through the wavelength-dependent refractive indices taken from \cite{ProcSPIE2017GSTnk}, as shown by Fig.~\ref{fig:GSTnk}.

\section{Results and discussion}
\subsection{Spectral tunability}
\begin{figure}[htbp]
\centering
\includegraphics[width=0.75\linewidth]{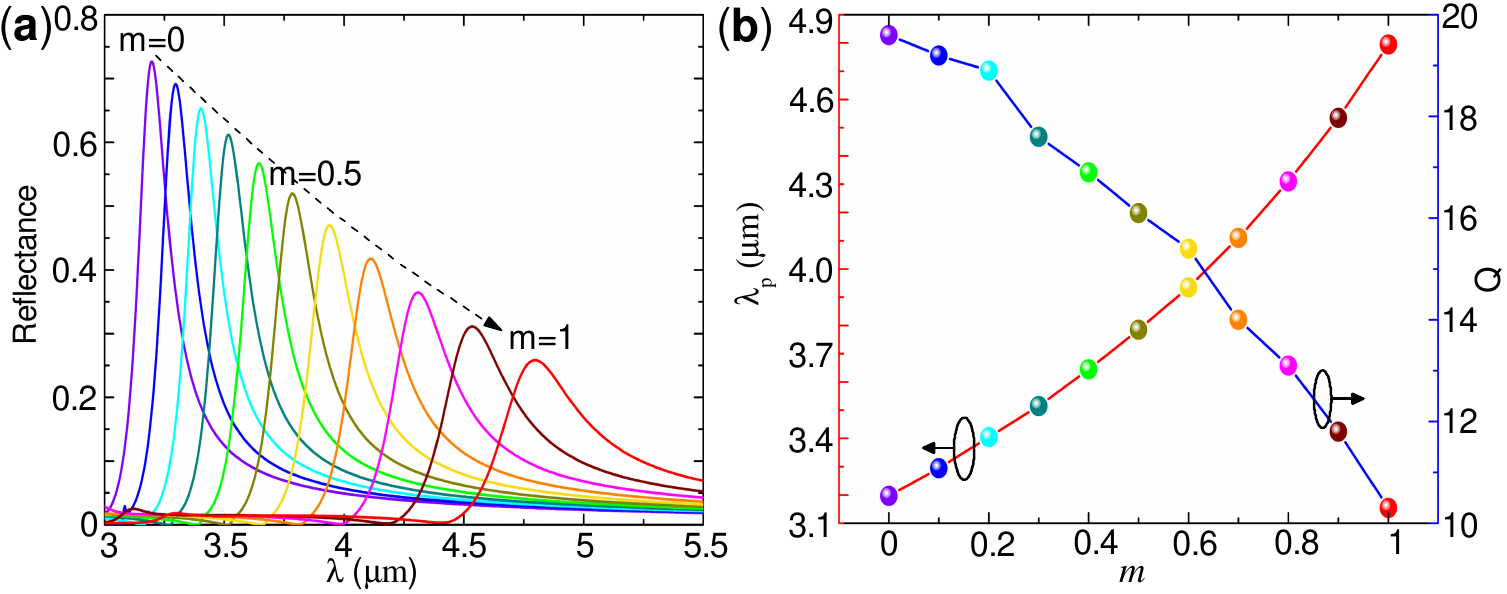}
\caption{(a) Simulated reflection spectra of the proposed filter for different GST crystallization fractions ranging from 0 (the amorphous state) to 1 (the crystalline state) in step of 0.1. (b) Peak reflectance wavelength and Q-factor as functions of GST crystallization fraction.}
\label{fig:Rspec}
\end{figure}

Figure~\ref{fig:Rspec}(a) shows the calculated reflection spectra of the proposed filter for different GST crystalline fractions. Remarkably, results show that as the GST crystallization fraction $m$ increases from 0 (the amorphous state) to 1 (the crystalline state), the reflection spectra are greatly red-shifted: the wavelength for the peak reflectance shifts from 3.197~$\mu$m to 4.795~$\mu$m. This corresponds to extremely large spectral tunability of 1.598~$\mu$m, or $2.13\Lambda$, which is slightly larger than the RA wavelength tunability of $\sim2\Lambda$. This striking spectral tunability is larger or even much larger than most of the reported GST-based mid-infrared filters, including those based on perfect absorbers \cite{JOSAB2013GstAbsTuneMIM,AM2015GstAbsTuneMIM,JOSAB2013GstAbsTuneMIM2,EPL2019GstAbsTuneMIM}, Fabry-P\'erot cavities \cite{AO2020GstFilterFP,OE2020GstFilterFP},  EOT effects \cite{AOM2016GstFilterEOT,Optica2020GstFilterEOT}, or SLRs combined with thin GST films \cite{OE2013GstFilterSLR,NL2013GstFilterSLR}. 

On the other hand, as the GST crystalline fraction increases from $m=0$ to 1, the peak reflectance decreases from 72.6\% to 25.8\%, the FWHM increases from 163~nm to 467~nm, and the corresponding Q-factor decreases from 19.6 to 10.3, as shown by Fig.~\ref{fig:Rspec}. The obtained Q-factors are larger than those based on perfect absorbers \cite{JOSAB2013GstAbsTuneMIM,AM2015GstAbsTuneMIM,JOSAB2013GstAbsTuneMIM2,EPL2019GstAbsTuneMIM}, and are comparable or larger than those based on EOT effects \cite{AOM2016GstFilterEOT,SPIE2018GstFilterEOT}.

In other words, compared with the literature, our design has extremely large spectral tunability, which covers the majority of the mid-infrared regime (3~$\mu$m to 5~$\mu$m). It also has relatively high reflection efficiency and meanwhile relatively large Q-factor.

\subsection{Physics mechanisms}
\begin{figure}[htbp]
\centering
\includegraphics[width=\linewidth]{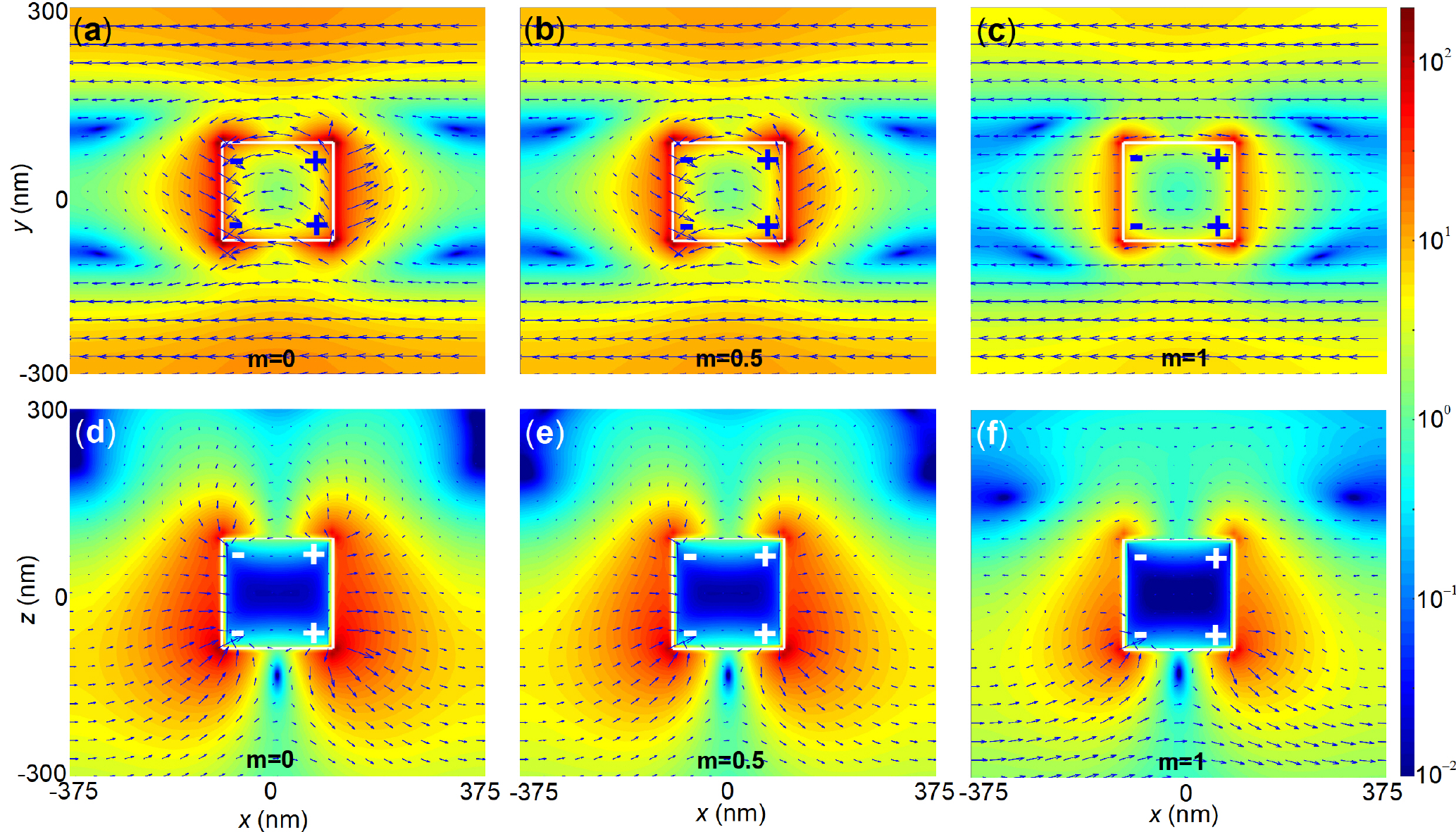}
\caption{Electric field intensity (in color) and vector (in arrows) maps at the three wavelengths of (a)(d) 3.197~$\mu$m, (b)(e) 3.785~$\mu$m, and (c)(f) 4.795~$\mu$m, corresponding to peak reflectance for $m=0$, 0.5 and 1, respectively.  (a)--(c) Top view at the bottom surface of gold nanorod ($z=-90$~nm), and (d)--(f) side view at $y=0$. The square-shaped gold nanorods surrounded by GST are outlined by white boxes. ``$+$'' and ``$-$'' indicate charge distributions.}
\label{fig:Fields}
\end{figure}

In order to understand the physics underlying the large spectral tunability and the high Q-factors of the designed filter, in Fig.~\ref{fig:Fields} we plot the calculated near-field electric field distributions at three wavelengths of 3.197~$\mu$m, 3.785~$\mu$m and 4.795~$\mu$m, which correspond to the peak reflectance for $m=0$, 0.5 and 1, respectively. For all these three wavelengths, results show that in-plane dipoles are excited in the gold nanorod and that the electric fields are greatly enhanced over a large volume, suggesting the excitation of in-plane SLRs. As the GST crystallization fraction increases from 0 to 1, the refractive index of GST increases dramatically. Therefore, the electric field enhancement slightly decreases, and the electric files are more tightly confined to the gold-GST interfaces, as shown by Fig.~\ref{fig:Fields}. This results in larger ohmic loss and thus smaller peak reflectance.

\begin{figure}[htbp]
\centering
\includegraphics[width=0.7\linewidth]{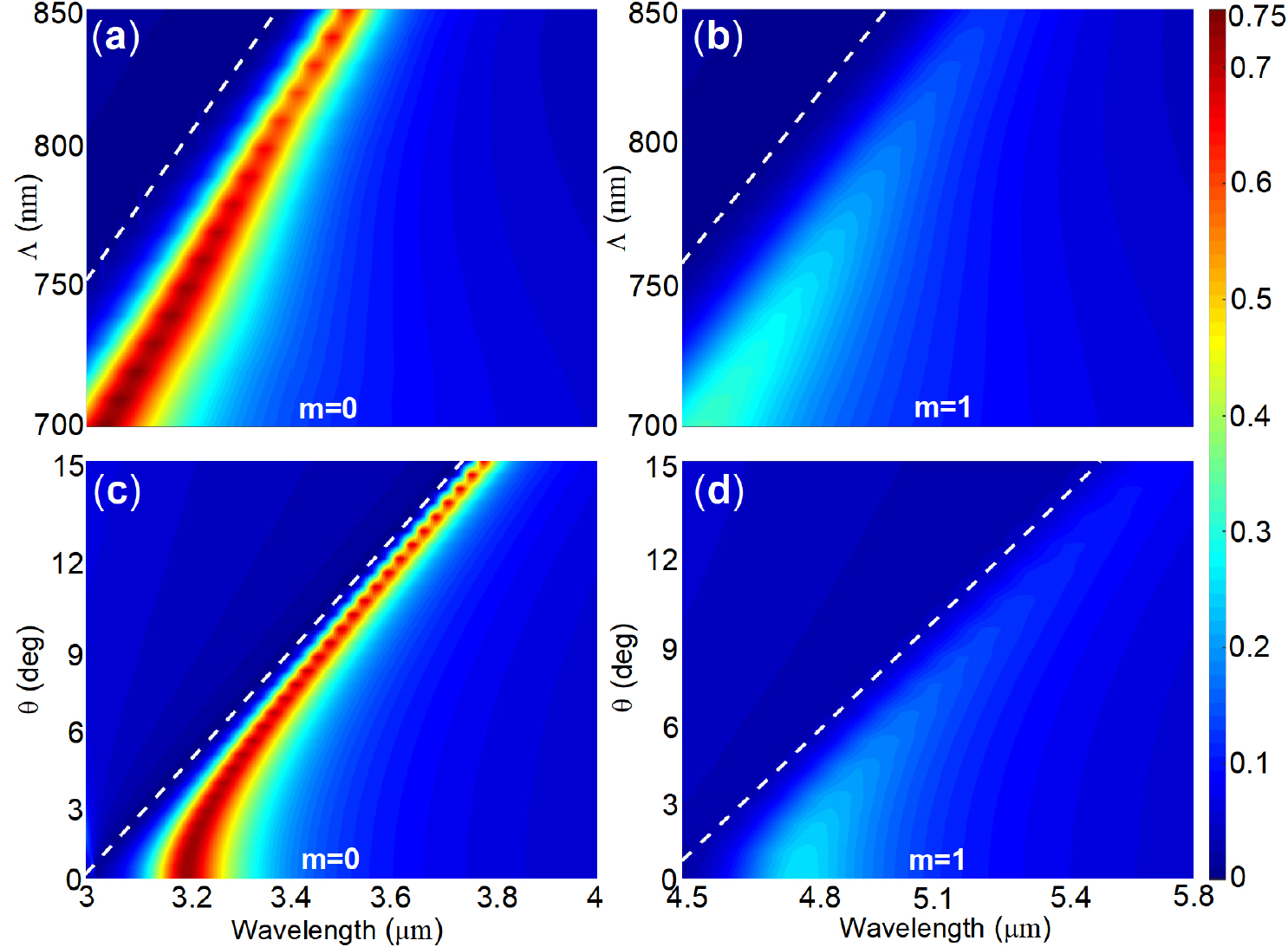}
\caption{Simulated reflection spectra of the proposed filter with GST in (a)(c) the amorphous state or (b)(d) the crystalline state as functions of (a)(b) the lattice period under normal incidence ($\theta=0$) and (c)(d) the incidence angle for $\Lambda=750$~nm. The white dashed curves represent RA wavelengths of the $(\pm1,0)$ order, which were calculated with Eq.~(\ref{Eq:RA}).}
\label{fig:Dispersion}
\end{figure}

In order to further validate the excitation of SLRs, we also calculated the reflection spectra of the proposed filter with GST in the amorphous state ($m=0$) or in the crystalline state ($m=1$) as functions of the lattice period $\Lambda$ and the incidence angle $\theta$. As $\Lambda$ increases from 700~nm to 850~nm, Figs.~\ref{fig:Dispersion}(a)(b) show that the peak reflectance decreases for both $m=0$ and $m=1$, and that the corresponding wavelengths increase linearly from 3.035~$\mu$m to 3.515~$\mu$m for $m=0$, and from 4.555~$\mu$m to 5.275~$\mu$m for $m=1$. Therefore, the spectral tuning ranges are 1.52~$\mu$m and 1.76~$\mu$m, corresponding to $2.17\Lambda$ and $2.07\Lambda$, respectively. Similarly, as $\theta$ increases from $0^\circ$ to $15^\circ$, Figs.~\ref{fig:Dispersion}(c)(d) also show that the peak reflectance decreases and that the corresponding wavelength increases for both $m=0$ and $m=1$. We note that the peak reflectance wavelengths generally follows the RA wavelengths of the $(\pm1,0)$ order, confirming the excitation of SLRs. Additionally, the FWHMs under oblique incidences are narrower than that under normal incidence, suggesting that the SLRs excited under oblique incidence have larger Q-factors. This is consistent with the fact out-of-plane SLRs excited under oblique incidences have larger Q-factors than in-plane SLRs excited under normal incidence \cite{OE2016Boyd_OLP}.

\subsection{Effects of gold nanorod sizes}
\begin{figure}[hbtp]
\centering
\includegraphics[width=0.98\linewidth]{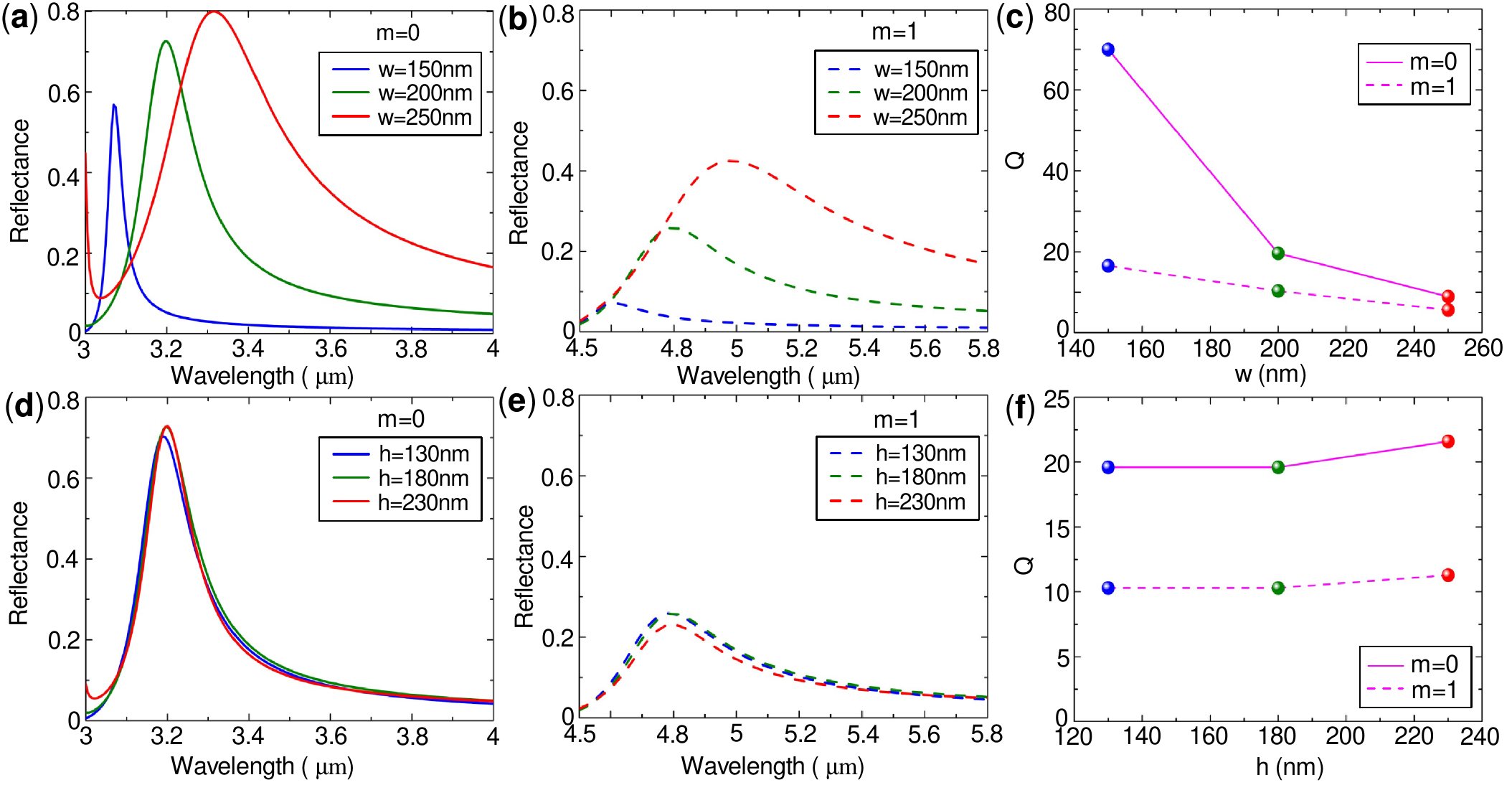}
\caption{Simulated reflection spectra of the proposed filter as functions of (a)(b) side length and (c)(d) thickness of gold nanorods. GST is in (a)(c) the amorphous state and (b)(d) the crystalline state. The calculations were performed with $\Lambda=750$~nm, $h=180$~nm for (a)(b), and $w=200$~nm for (c)(d).}
\label{fig:SizeEffect}
\end{figure}

We now study the effects of the gold nanorod sizes on the filtering performance. Figures~\ref{fig:SizeEffect}(a)(b) shows that if the side length of gold nanorods $w$ decreases from 200~nm (blue curves) to 150~nm (purple curves), the peak reflectance decreases slightly from 72.6\% to 56.8\% for GST in the amorphous state ($m=0$), but decreases dramatically from 25.8\% to 7.1\% for GST in the crystalline state ($m=1$). At the cost of decreased reflectance, the corresponding Q-factors increase tremendously from 19.6 to 70 for $m=0$, and from 10.3 to 16.5 for $m=1$, as shown in Fig.~\ref{fig:SizeEffect}(c). If $w$ increases to 250~nm (red curves), the peak reflectance further increases from 72.6\% to 80.0\% (or from 25.8\% to 42.5\%) at the cost that the Q-factor decreases from 19.6 to 8.9 (or from 10.3 to 5.6) for $m=0$ (or $m=1$). Therefore, by carefully designing the gold nanorod side length, the peak reflectance can be greatly increased at the cost of relatively smaller Q-factor or the Q-factor can be significantly improved at the cost of relative smaller peak reflectance. This adjustable filtering performance makes our design very attractive because it can meet various demands.

If the thickness of gold nanorods $h$ increases or decreases by the same amount of 50~nm, however, Figs.~\ref{fig:SizeEffect}(d)--(f) show that the peak reflectance as well as the Q-factor have negligible variations for both $m=0$ and $m=1$. In other words, $h$ has little influences on the filtering performance. Such extremely large tolerance on the gold nanorod thickness will greatly facilitate the fabrication of the proposed filter.

\section{Conclusions}
In conclusions, we have proposed a nonvolatile, reconfigurable, and narrowband mid-infrared bandpass filter based on SLRs in phase-change material GST. Results have shown that when GST transits from the amorphous state ($m=0$) to the crystalline state ($m=1$), the narrowband reflection spectrum of the proposed filter can tuned from $3.197~\mu$m to $4.795~\mu$m, covering the vast majority of the mid-infrared 3~$\mu$m to 5~$\mu$m atmospheric window. For $m=0$ the peak reflectance reaches 72.6\% and the Q-factor is up to 19.3, while for $m=1$ the peak reflectance reaches 25.8\% and the Q-factor is 10.3. Near-field distributions, and the strong dependence of the reflection spectra on the lattice period and the incidence angle have confirmed that the narrowband filtering originates from the excitation of SLR. We have also found that the Q-factor can be greatly increased to $Q=70$ for $m=0$ (or $Q=16.5$ for $m=1$) at the cost of relatively smaller peak reflectance, and the peak reflectance can be improved to 80.0\% for $m=0$ (or 42.5\% for $m=1$) at the cost of relatively smaller Q-factors. We expect that the designed nonvolatile reconfigurable narrowband mid-infrared filters will find applications in chemical spectroscopy, thermography, and multispectral/hyperspectral imaging. We also expect that dynamically tunable SLRs based on phase change materials can be extended to other spectral regimes and to other applications besides filtering.

\section*{Acknowledgments}
This work was supported by the Shenzhen Fundamental Research and Discipline Layout project (JCYJ20180507182444250), and by the State Key Laboratory of Advanced Optical Communication Systems and Networks, China (No. 2019GZKF2).

\bibliographystyle{unsrt}
\bibliography{sample}

\begin{thebibliography}{10}

\bibitem{OLE2006Monitoring}
U.~Willer, M.~Saraji, A.~Khorsandi, P.~Geiser, and W.~Schade.
\newblock Near-and mid-infrared laser monitoring of industrial processes,
  environment and security applications.
\newblock {\em Opt. Lasers Eng.}, 44:699--710, 2006.

\bibitem{ApplSpec2012Spec}
5.~R. Bhargava.
\newblock Infrared spectroscopic imaging: the next generation.
\newblock {\em Appl. Spectrosc.}, 66:1091--1120, 2012.

\bibitem{CBP2016thermography}
G.~J. Tattersall.
\newblock Infrared thermography: A non-invasive window into thermal physiology.
\newblock {\em Comp. Biochem. Physiol., Part A: Mol. Integr. Physiol.},
  202:78--98, 2016.

\bibitem{ProcSPIE2018MidIRfilterAO}
J.~Ward and S.~Valle.
\newblock Acousto-optic devices for operation in the infrared.
\newblock {\em Proc. SPIE}, 10683:1068327, 2018.

\bibitem{AO2018MidIRfilterAO}
Oleg~I. Korablev, Denis~A. Belyaev, Yuri~S. Dobrolenskiy, Alexander~Y.
  Trokhimovskiy, and Yuri~K. Kalinnikov.
\newblock Acousto-optic tunable filter spectrometers in space missions
  {[Invited]}.
\newblock {\em Appl. Opt.}, 57:C103--C119, 2018.

\bibitem{ProcSPIE2016MidIRfilterMEMS}
M.~Ebermann, N.~Neumann, K.~Hiller, M.~Seifert, M.~Meinig, and S.~Kurth.
\newblock Tunable {MEMS Fabry-P\'erot} filters for infrared microspectrometers:
  a review.
\newblock {\em Proc. SPIE}, 9760:97600H, 2016.

\bibitem{JMM2020MidIRfilterLC}
Ziyu Wang, Jasleen~Kaur Lall, Yannick Folwill, and Hans Zappe.
\newblock Compact tunable mid-infrared {Fabry–P\'erot} filters actuated by
  liquid crystal elastomers.
\newblock {\em J. Micromech. Microeng.}, 30:075002, 2020.

\bibitem{Optica2020GstFilterEOT}
Matthew~N. Julian, Calum Williams, Stephen Borg, Scott Bartram, and Hyun~Jung
  Kim.
\newblock Reversible optical tuning of {GeSbTe} phase-change metasurface
  spectral filters for mid-wave infrared imaging.
\newblock {\em Optica}, 7:746--754, 2020.

\bibitem{NatPhoton2017PCMrev}
M.~Wuttig, H.~Bhaskaran, and T.~Taubner.
\newblock Phase-change materials for non-volatile photonic applications.
\newblock {\em Nat. Photon.}, 11:465--476, 2017.

\bibitem{ATS2019PCMrev}
Tun Cao and Mengjia Cen.
\newblock Fundamentals and applications of chalcogenide phase-change material
  photonics.
\newblock {\em Adv. Theory Simul.}, 2:1900094, 2019.

\bibitem{AOM2019PCMmsRev}
Fei Ding, Yuanqing Yang, and Sergey~I. Bozhevolnyi.
\newblock Dynamic metasurfaces using phase-change chalcogenides.
\newblock {\em Adv. Optical Mater.}, 7:1801709, 2019.

\bibitem{NanoP2020PCMrev}
Sajjad Abdollahramezani, Omid Hemmatyar, Hossein Taghinejad, Alex Krasnok,
  Yashar Kiarashinejad, Mohammadreza Zandehshahvar, Andrea Alù, and Ali Adibi.
\newblock Tunable nanophotonics enabled by chalcogenide phase-change materials.
\newblock {\em Nanophotonics}, 9:20200039, 2020.

\bibitem{JOSAB2013GstAbsTuneMIM}
Tun Cao, Lei Zhang, Robert~E. Simpson, and Martin~J. Cryan.
\newblock Mid-infrared tunable polarization-independent perfect absorber using
  a phase-change metamaterial.
\newblock {\em J. Opt. Soc. Am. B}, 30:1580--1585, 2013.

\bibitem{JOSAB2013GstAbsTuneMIM2}
Tun Cao, Robert~E. Simpson, and Martin~J. Cryan.
\newblock Study of tunable negative index metamaterials based on phase-change
  materials.
\newblock {\em J. Opt. Soc. Am. B}, 30:439--444, 2013.

\bibitem{AM2015GstAbsTuneMIM}
Andreas Tittl, Ann-Katrin~U. Michel, Martin Sch{\"a}ferling, Xinghui Yin,
  Behrad Gholipour, Long Cui, Matthias Wuttig, Thomas Taubner, Frank Neubrech,
  and Harald Giessen.
\newblock A switchable mid-infrared plasmonic perfect absorber with
  multispectral thermal imaging capability.
\newblock {\em Adv. Mater.}, 27:4597--4603, 2015.

\bibitem{EPL2019GstAbsTuneMIM}
Ximin Tian, Junwei Xu, Kun Xu, Xiaolong Ma, Xiangyang Duan, Peng Yang, Pei
  Ding, and Zhi-Yuan Li.
\newblock Wavelength-selective, tunable and switchable plasmonic perfect
  absorbers based on phase change materials {Ge$_2$Sb$_2$Te$_5$}.
\newblock {\em EPL}, 128:67001, 2019.

\bibitem{SM2019GstFilterRg}
Xiya Ding, Xiao Yang, Jingjing Wang, Kai Guo, Fei Shen, Hongping Zhou, Rui Sun,
  Zhizhong Ding, Jun Gao, and Zhongyi Guo.
\newblock Theoretical analysis and simulation of a tunable mid-infrared filter
  based on {Ge2Sb2Te5 (GST)} metasurface.
\newblock {\em Superlatt. Microstruct.}, 132:106169, 2019.

\bibitem{AO2020GstFilterFP}
Kun Zhou, Qiang Cheng, Lu~Lu, Bowen Li, Jinlin Song, Mengting Si, and Zixue
  Luo.
\newblock Multichannel tunable narrowband mid-infrared optical filter based on
  phase-change material {Ge$_2$Sb$_2$Te$_5$} defect layers.
\newblock {\em Appl. Opt.}, 59:595--600, 2020.

\bibitem{OE2020GstFilterFP}
Calum Williams, Nina Hong, Matthew Julian, Stephen Borg, and Hyun~Jung Kim.
\newblock Tunable mid-wave infrared {Fabry-Perot} bandpass filters using
  phase-change {GeSbTe}.
\newblock {\em Opt. Express}, 28:10583--10594, 2020.

\bibitem{AOM2016GstFilterEOT}
Miquel Rud\'e, Vahagn Mkhitaryan, Arif~Engin Cetin, Timothy~Alan Miller, Albert
  Carrilero, Simon Wall, Francisco Javier~García de~Abajo, Hatice Altug, and
  Valerio Pruneri.
\newblock Ultrafast and broadband tuning of resonant optical nanostructures
  using phase-change materials.
\newblock {\em Adv. Optical Mater.}, 4:1060--1066, 2016.

\bibitem{SPIE2018GstFilterEOT}
L.~Trimby, D.~Wright, and A.~Baldycheva.
\newblock Phase-change band-pass filters for multispectral imaging.
\newblock {\em Proc. SPIE}, 10541:105412B, 2018.

\bibitem{ChemRev2018GrigorenkoSLRreview}
V.~G. Kravets, A.~V. Kabashin, W.~L. Barnes, and A.~N. Grigorenko.
\newblock Plasmonic surface lattice resonances: A review of properties and
  applications.
\newblock {\em Chem. Rev.}, 118:5912--5951, 2018.

\bibitem{NanoRes2018ZhengSLRreview}
B.~B. Rajeeva, L.~Lin, and Y.~Zheng.
\newblock Design and applications of lattice plasmon resonances.
\newblock {\em Nano Res.}, 11:4423--4440, 2018.

\bibitem{MatToday2018OdomSLRreview}
W.~Wang, M.~Ramezani, A.~I. V{\"a}kev{\"a}inen, P.~T{\"o}rm{\"a}, J.~G. Rivas,
  and T.~W. Odom.
\newblock The rich photonic world of plasmonic nanoparticle arrays.
\newblock {\em Mater. Today}, 21:303--314, 2018.

\bibitem{ACSNano2014SLRfilter}
Zhongyang Li, Serkan Butun, and Koray Aydin.
\newblock Ultranarrow band absorbers based on surface lattice resonances in
  nanostructured metal surfaces.
\newblock {\em ACS Nano}, 8:8242--8248, 2014.

\bibitem{OE2013GstFilterSLR}
Y.~G. Chen, T.~S. Kao, B.~Ng, X.~Li, X.~G. Luo, B.~Luk'yanchuk, S.~A. Maier,
  and M.~H. Hong.
\newblock Hybrid phase-change plasmonic crystals for active tuning of lattice
  resonances.
\newblock {\em Opt. Express}, 21:13691--13698, 2013.

\bibitem{NL2013GstFilterSLR}
Ann-Katrin~U. Michel, Dmitry~N. Chigrin, Tobias W.~W. Ma\ss, Martin~Saling
  Kathrin Sch\"onauer~and, Matthias Wuttig, and Thomas Taubner.
\newblock Using low-loss phase-change materials for mid-infrared antenna
  resonance tuning.
\newblock {\em Nano Lett.}, 13:3470--3475, 2013.

\bibitem{ProcSPIE2017GSTnk}
Li~Tian Chew, Weiling Dong, Li~Liu, Xilin Zhou, Jitendra Behera, Hailong Liu,
  Kandammathe~V. Sreekanth, Libang Mao, Tun Cao, Joel Yang, and Robert~E.
  Simpson.
\newblock Chalcogenide active photonics.
\newblock {\em Proc. SPIE}, 10345:103451B, 2017.

\bibitem{JOSAA1995RCWA}
M.~G. Moharam, Drew~A. Pommet, Eric~B. Grann, and T.~K. Gaylord.
\newblock Stable implementation of the rigorous coupled-wave analysis for
  surface-relief gratings: Enhanced transmittance matrix approach.
\newblock {\em J. Opt. Soc. Am. A}, 12(5):1077--1086, 1995.

\bibitem{JOSAA1997RCWA}
P.~Lalanne.
\newblock Improved formulation of the coupled-wave method for two-dimensional
  gratings.
\newblock {\em J. Opt. Soc. Am. A}, 14:1592--1598, 1997.

\bibitem{PRB2006RCWA}
A.~David and H.~Benisty.
\newblock Fast factorization rule and plane-wave expansion method for
  two-dimensional photonic crystals with arbitrary hole-shape.
\newblock {\em Phys. Rev. B: Condens. Matter Mater.}, 73:075107, 2006.

\bibitem{Au2012PRB}
R.~L. Olmon, B.~Slovick, T.~W. Johnson, D.~Shelton, S.-H. Oh, G.~D. Boreman,
  and M.~B. Raschke.
\newblock Optical dielectric function of gold.
\newblock {\em Phys. Rev. B: Condens. Matter Mater.}, 86:235147, 2012.

\bibitem{AJP1982LLmodel}
D.~Aspnes.
\newblock Local-field effects and effective-medium theory: a microscopic
  perspective.
\newblock {\em Am. J. Phys.}, 50:704--709, 1982.

\bibitem{OE2016Boyd_OLP}
M.~J. Huttunen, Ksenia Dolgaleva, P.~T{\"o}rm{\"a}, and Robert~W. Boyd.
\newblock Ultra-strong polarization dependence of surface lattice resonances
  with out-of-plane plasmon oscillations.
\newblock {\em Opt. Express}, 24:28279--28289, 2016.

\end{thebibliography}

\end{document}